\documentclass[]{aastex631}
\submitjournal{Astronomical Journal, Accepted version}

\newcommand{\f}{\textit{Firmamento}}
\newcommand{\fb}{\textbf{\textit{Firmamento}}}
\newcommand{\gr}{$\gamma$-ray}

\newcommand{\nup}{$\nu_{\rm peak}$}
\newcommand{\lsim}{{\lower.5ex\hbox{$\; \buildrel < \over \sim \;$}}}
\newcommand{\gsim}{{\lower.5ex\hbox{$\; \buildrel > \over \sim \;$}}}
\newcommand{\nufnu}{$\nu$f$({\nu})$}

\begin{document}

\title{\fb: a multi-messenger astronomy tool for citizen and professional scientists}

\author{Dhurba Tripathi}
\affiliation{New York University Abu Dhabi, 
PO Box 129188 Abu Dhabi \\ United Arab Emirates}
\author[0000-0002-2265-5003]{Paolo Giommi}
\affiliation{Center for Astrophysics and Space Science (CASS),
New York University Abu Dhabi\\ 
PO Box 129188 Abu Dhabi\\ United Arab Emirates}
\affiliation{Institute for Advanced Study -
Technische Universit{\"a}t M{\"u}nchen\\
Lichtenbergstrasse 2a, D-85748 Garching bei M\"unchen, Germany}
\affiliation{Associated to INAF, 
Osservatorio Astronomico di Brera\\
via Brera, 28, I-20121 Milano, Italy}
\author[0000-0002-8462-4894]{Adriano Di Giovanni}
\affiliation{Gran Sasso Science Institute,
I-67100 L'Aquila AQ, Italy}
\affiliation{Center for Astrophysics and Space Science (CASS),
New York University Abu Dhabi\\ 
PO Box 129188 Abu Dhabi\\ United Arab Emirates}
\author{Rawdha R. Almansoori}
\affiliation{New York University Abu Dhabi, 
PO Box 129188 Abu Dhabi \\ United Arab Emirates}
\author{Nouf Al Hamly} 
\affiliation{New York University Abu Dhabi, 
PO Box 129188 Abu Dhabi \\ United Arab Emirates}
\author[0000-0002-1061-0510]{Francesco Arneodo}
\affiliation{New York University Abu Dhabi, 
PO Box 129188 Abu Dhabi \\ United Arab Emirates}
\affiliation{Center for Astrophysics and Space Science (CASS),
New York University Abu Dhabi\\ 
PO Box 129188 Abu Dhabi\\ United Arab Emirates}
\author[0000-0002-8171-6507]{Andrea V. Macci\`o}
\affiliation{New York University Abu Dhabi, 
PO Box 129188 Abu Dhabi \\ United Arab Emirates}
\affiliation{Center for Astrophysics and Space Science (CASS),
New York University Abu Dhabi\\ 
PO Box 129188 Abu Dhabi\\ United Arab Emirates}
\affiliation{Max-Planck-Institut f\"ur Astronomie, \\
K\"onigstuhl 17, 69117 Heidelberg, Germany}
\author[0000-0001-5427-5155]{Goffredo Puccetti}
\affiliation{New York University Abu Dhabi, 
PO Box 129188 Abu Dhabi \\ United Arab Emirates}
\author[0000-0001-7909-588X]{Ulisses Barres de Almeida}
\affiliation{Centro Brasileiro de Pesquisas F\'isicas,\\
Rua Dr. Xavier Sigaud 150, 22290-180, Rio de Janeiro, Brazil}
\author[0000-0001-6679-3777]{Carlos Brandt}
\affiliation{Constructor University,\\ 
Campus Ring 1, D-28759 Bremen, Germany}
\author[0000-0002-9744-071X]{Simonetta Di Pippo}
\affiliation{SDA Bocconi,\\ 
Via Sarfatti 25, I-20100, Milano, Italy}
\affiliation{New York University Abu Dhabi, 
PO Box 129188 Abu Dhabi \\ United Arab Emirates}
\author[0000-0001-9104-3214]{Michele Doro}
\affiliation{University of Padova,\\
via Marzolo 8, I-35131 Padova (Italy)}
\author[0000-0002-5804-6605]{Davit Israyelyan}
\affiliation{ICRANet-Armenia\\
Marshall Baghramian Avenue 24a, Yerevan 0019, Armenia}
\author[0000-0002-6737-538X]{A.M.T. Pollock}
\affiliation{Department of Physics and Astronomy,\\ 
University of Sheffield, Sheffield S3 7RH, UK}
\author[0000-0003-2011-2731]{Narek Sahakyan}
\affiliation{ICRANet-Armenia\\
Marshall Baghramian Avenue 24a, Yerevan 0019, Armenia}







\begin{abstract}
\f\ (https://firmamento.hosting.nyu.edu) is a new-concept web-based and mobile-friendly data analysis tool dedicated to multi-frequency/multi-messenger emitters, as exemplified by blazars. 
Although initially intended to support a citizen researcher project at New York University-Abu Dhabi (NYUAD), \f\, has evolved to be a valuable tool for professional researchers due to its broad accessibility to classical and contemporary multi-frequency open data sets. 
From this perspective \f\ facilitates the identification of new blazars and other multi-frequency emitters 
in the localisation uncertainty regions of sources detected by current and planned observatories such as  Fermi-LAT, Swift , eROSITA,  CTA,  ASTRI Mini-Array, LHAASO, IceCube, KM3Net, SWGO, etc. The multi-epoch and multi-wavelength data that \f\, retrieves from over  90 remote and local catalogues and databases can  be used to characterise the spectral energy distribution and the variability properties of cosmic sources as well as to constrain physical models. 
\f\, distinguishes itself from other online platforms due to its high specialization, the use of machine learning and other methodologies to characterise the data and for its commitment to inclusivity. From this particular perspective, its objective is to assist both researchers and citizens interested in science, strengthening a trend that is bound to gain momentum in the coming years as data retrieval facilities improve in power and machine learning/artificial intelligence tools become more widely available.
\end{abstract}




\section{Introduction} \label{sec:intro}

The increasing trend in astronomy facilities to provide unrestricted access to their data at some stage, coupled with the push for the adoption of FAIR principles 
\footnote{https://www.go-fair.org/fair-principles},  together with the presence of projects such as the well-established International Virtual Observatory Alliance\footnote{https://ivoa.net}, and many others that have emerged in recent years are significantly expanding the discovery potential in the scientific community.
Initiatives of this type include the European Science Cloud (EOSC)\footnote{https://www.eosc-portal.eu}, the Research Data Alliance (RDA)\footnote{https://www.rd-alliance.org}, the Research Infrastructure Cluster (ASTERICS)\footnote{https://www.asterics2020.eu/}, and Open Universe\footnote{https://openuniverse.asi.it}, among others. 
In addition, the rapidly increasing number of online data archives, funded by most space agencies and other organizations, provide a wide array of open-access astronomical data. Listing all the currently operating data services would be impractical due to their abundance, a situation that underscores the rapid expansion of the data available and, consequently, the discovery potential within the scientific community. 
Many of the principles and part of the software outlined in this paper stem from the efforts undertaken over the past few years within the Open Universe initiative \citep{GiommiOU}. This international collaboration was proposed at the United Nations Committee on the Peaceful Uses of Outer Space (COPUOS) in 2016 and has been extensively discussed in various meetings coordinated by the United Nations Office for Outer Space Affairs (UNOOSA).

These positive developments are also creating unprecedented opportunities for individuals with diverse skills to participate in scientific activities and discovery.
 Along these lines, there are important examples of successful projects involving large numbers of citizen scientists, often without formal scientific training in the field of astronomy. These include Galaxy Zoo \footnote{http://zoo1.galaxyzoo.org/}, which led to the identification of thousands of galaxies based on their shape, 
and SETI@home \footnote{https://setiathome.berkeley.edu/}, which involve citizens with Internet-connected computers in the Search for Extraterrestrial Intelligence (SETI). 
Many scientific papers involving citizen scientists have already appeared in the professional literature demonstrating that also untrained individuals can contribute in a significant way to research. 
Current citizen scientists initiatives are based on scientific projects that are carefully defined by professional scientists, while citizen scientists, often in large numbers, contribute by providing the manpower that is needed to carry out predefined tasks. 
The rapid proliferation of machine learning tools and freely accessible artificial intelligence (AI) services is significantly accelerating the path towards the ultimate objective of enabling normal citizens with an interest in science to autonomously conceive and carry-out research projects, one of the long-term goals of the Open Universe initiative.

Despite the positive trend towards unrestricted access to data, challenges still persist as the complete utilization of all available datasets often requires analysis of data from vastly different detectors, each demanding specialized expertise that is nearly impossible to meet. However, the increasing availability of calibrated and science-ready multi-frequency data is making multi-messenger astronomy accessible to a wider range of researchers and citizen scientists.

This paper introduces \f, a new online tool developed to support citizen and professional researchers in the field of blazars and multi-messenger research, building upon the experience of the Open Universe initiative. Accessible from a computer, tablet, or mobile phone, this service is a powerful instrument for identifying and studying cosmic sources detected by current and upcoming multi-messenger detectors with localisation uncertainties ranging from a few arc-seconds to several square degrees including, but not limited to, eRosita \citep{erosita2020}, Swift \citep{swift}, Fermi \citep{Fermi-LAT}, AGILE \citep{AGILE}, LHAASO \citep{LHAASO}, CTA \citep{CTA}, ASTRI Mini-Array \citep{Astri-Mini-Array}, IceCubeGen2 \citep{IceCubeGen2}, KM3Net \citep{KM3NET}. Due to its user-friendly interface, even individuals without astronomical expertise can utilize \f, making it a useful tool also in educational contexts.

\section{Multi-messenger astrophysics}

In parallel with the achievements of the long-established field of multi-frequency astronomy, the recent development of highly advanced observatories capable of detecting other cosmic messengers such as neutrinos \footnote{e.g. https://icecube.wisc.edu}, ultra-high-energy cosmic rays \footnote{e.g. https://www.auger.org} 
and gravitational waves \footnote{e.g. https://www.ligo.caltech.edu }, paved the way for multi-messenger astrophysics. 
This new development is  rapidly widening our knowledge of the Universe through the study of matter and radiation in astrophysical contexts where the gravitational, electromagnetic, strong and weak forces are at work in an extremely wide range of physical conditions. 
Neutrinos, in particular, provide a unique observational window on the most extreme astrophysical environments. That is because these weakly interacting particles are stable, can travel cosmological distances and, unlike photons, can escape from the dense regions where they are produced and can reach the Earth unattenuated even at the highest energies.  

Despite a flux of neutrinos and of high-energy cosmic rays has been detected \citep{IceCube2013}, the nature of the sources that emit these particles largely remains an unresolved issue.  Different types of astrophysical sources, located both in our Galaxy and outside, have been proposed to be multi-messenger emitters. Among these, the class of blazars appears to be among the most promising.    

\subsection{Blazars}

Blazars are the most powerful persistent sources in the Universe \citep{Urry1995}.
They are a rare type of highly-variable Active Galactic Nuclei \citep[AGN,][]{AGNReview} that emit electromagnetic radiation over the entire energy spectrum, from radio waves to high-energy gamma-rays, and are suspected to be the sources of high-energy neutrinos \citep{Aartsen2018,GiommiPadovani2021}. The physical engine that powers these remarkable sources is thought to be a relativistic jet of matter that moves away from the central supermassive black hole in directions that happen to be aligned toward the Earth. This very special physical and geometrical condition is what make blazars so peculiar and cause them to be detected at energies where other types of cosmic emitters are not observed, like in the \gr\, band where they are by far the most common type of extragalactic sources observed \citep{4FGL,2019A&A...627A..13B}. Despite that, only about 6,000 such objects have been so far discovered in the many radio, X-ray and \gr\, surveys that have been carried out over the last 40 years \citep{GiommiPadovani2021}. With the rapidly growing availability of new data, this number is bound to increase significantly very soon. For instance the on-going SRG/eRosita X-ray sky survey \citep{erosita2020,efeds} is expected to detect well over 100,000 blazars. The first large catalogue from this survey has been announced to be published in the coming  months.
A smaller but still important number of blazars will be detected by upcoming observatories operating in the very high-energy gamma-ray band, like CTA\footnote{\url{https://www.cta-observatory.org}}\citep{CTA},
ASTRI Mini-Array\footnote{http://astri.me.oa-brera.inaf.it/en/}\citep{ASTRI-MiniArray}, SWGO \citep{SWGO} and by other multi-messenger facilities like KM3NeT\footnote{\url{https://www.km3net.org}}\citep{KM3NET}, Baikal-GVD\footnote{\url{https://baikalgvd.jinr.ru}} \citep{Baikal}, and the Pacific Ocean Neutrino Experiment \citep[P-ONE,][]{PONE}.
Identifying and characterising all these sources will be a challenging task. \f\, has the potential for playing a significant role in this process, also through citizen scientists and students projects.  

\section{\f\,: a data analysis tool for blazars discovery and multi-messenger research}
\f , in its present version (https://firmamento.hosting.nyu.edu), focuses on blazar discovery and on multi-messenger astrophysics research.
To do so, it provides:
\begin{enumerate}
\item	localisation uncertainty maps useful for the identification of the counterparts of X-ray, \gr\,, high-energy neutrino, and any other type of astronomical sources with non-negligible positional errors;
\item	Spectral Energy Distributions (SEDs) obtained via VOU-Blazars \citep{VOU-Blazars}, V2.00, a tool that retrieves, homogenises and combines data from over 90 catalogues and spectral databases. SEDs are then generated and plotted after de-reddening and converting all measurements to monochromatic \nufnu\ fluxes in common units (in frequency and intensity).
\item access to several classical and recently released astronomical surveys via an adapted version of the aladin \footnote{\url{https://aladin.u-strasbg.fr}} sky visualiser that has been integrated into the \f\, tool.
Imaging data in most energy bands of the electromagnetic spectrum, e.g. radio, infrared, optical, UV, X-ray and gamma-rays, are provided;
\item machine learning and other software suitable for the characterisation of blazar broad-band spectra;
\item links to other astronomical sites and space news;
\item documentation and (video) tutorials.
\end{enumerate}

\subsection{\f's architecture}

\begin{figure}
\centering \includegraphics[width=15.cm]{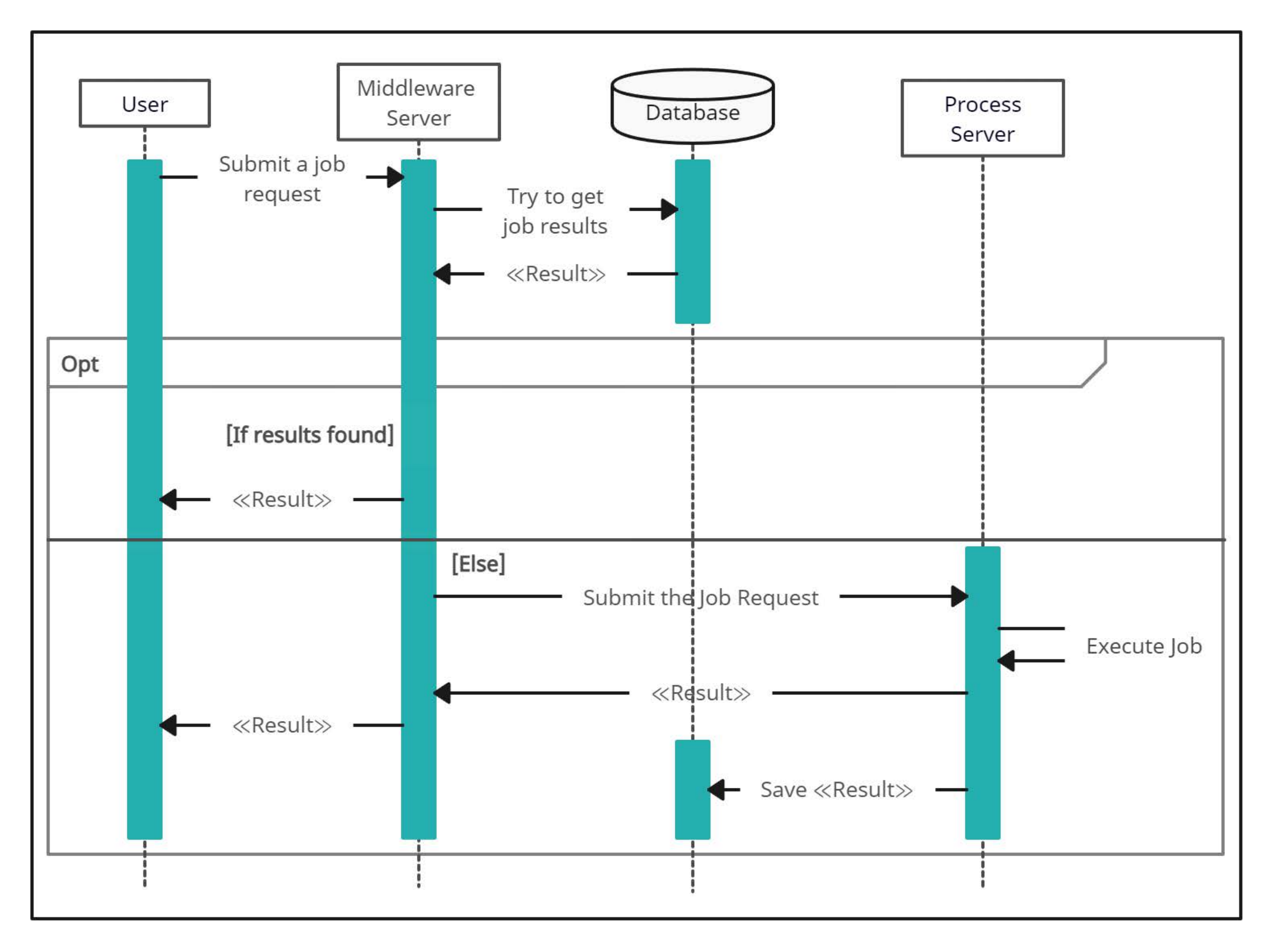}
\caption{A schematic view of \f's architecture.
\label{architecture}}
\end{figure}  

Fig. \ref{architecture} illustrates the architecture of the \f\ tool, which includes four main components: a front-end web interface, a middleware server, a database, and a processing server. Each component plays a crucial role in the overall functionality of the system.
In the following sections, we give a description of each component, as well as of the software running on the server.

\subsubsection{The front-end Web pages} The front-end website serves as the primary user interface for most online services. In a system like \f\,, which aims at serving users with varying skill levels, this component is a particularly important. Our design follows a minimalistic approach with a focus on ease of use. 
The site includes four main areas that can be accessed from dedicated tabs. 
\begin{itemize}
\item  {\bf Home}, which provides an introduction to the site and general information
\item  {\bf Data access}, where the multi-messenger data of astrophysical sources can be requested, retrieved, plotted and analysed, 
\item  {\bf Resources} provides links to a number of external sites about astronomy and space science 
\item {\bf Media} is where users can browse through media sites about space news, streaming space TVs, NASA picture of the day, and image galleries.
\item {\bf Tutorials} for video tutorials and documentation
\item {\bf Feedback} is a page dedicated to user feedback.
\end{itemize}
From a technical viewpoint the front-end of \f, was built using the React JavaScript framework \footnote{\url{https://reactjs.org/}}, taking special care to ensure that it can be easily used from both computers and smart phones.

\subsubsection{The middleware server}
 The middleware server is hosted on a cloud service provider, presently Heroku but subject to change. It serves as a crucial component in our system, acting as a gateway to the NYUAD server and, relying on the security measures provided by the cloud service. Its primary functions encompass establishing rate limits, validating user inputs, and implementing safety measures to safeguard the process server at NYUAD against potential cyber threats. 

Its operational workflow involves the middleware server conducting preliminary checks in the database when a user requests a new job. If the relevant data is not found, the request is then forwarded to the process server at NYUAD, with the middleware server actively monitoring the process server's status. A noteworthy aspect is the utilization of websockets for communication, facilitating real-time logs visible on the front-end during job execution. This server also serves as a queue manager for incoming jobs, because the current system can only execute a small number of jobs at the one time. Additionally, its role in job distribution contributes to the system's scalability. As the number of process servers increases, the middleware server seamlessly functions as a load balancer, distributing jobs among different process servers, optimizing resource utilization.  
\subsubsection{The cloud-based data repository}
A cloud-based database service (currently firebase, and subject to change) is used to store the results of the jobs that have been executed by the process server. This database can only be accessed by the middleware server and the process server. Access from other locations is blocked thus making it secure. Whenever an incoming request comes to the middleware server, the server checks this database if the job was previously done as it has database access rights. In case if results for the job are not present, a null value is sent to the middleware server signalling a need to contact the NYUAD process server.
\subsubsection{The server @ NYUAD}
The server located at NYUAD is where the requests by the users are directed for appropriate processing, when necessary. This crucial component of the firmamento system is a server shared with other NYUAD projects. When a user initiates a new job, the request passes through the middleware component, to this server, signalling that the job has not been processed before. This server currently can run a maximum of five jobs at the same time as these may be demanding in terms of available resources. A queue system is therefore established to prioritise jobs on a first come first served basis. During the execution of the job, all the logs are sent to the middleware server through web sockets, thus making it possible for realtime logs. Once a job is executed successfully, results are submitted to the middleware server which serves the results to the front end. 
Results include a SED graph, a CSV file including the SED data, two text files, and the output of the \textit{Blast} software. Once the results are available on the middleware server, the user is able to view them on the website. In case of any errors, users are notified on the front-end and an error log is kept on the server to be analysed by the developers for possible recovering actions.

The server at NYUAD is also highly secure as communication to the server is only possible through middleware server and the cloud database. Any other attempted connection to the server is blocked by our security system, which was developed in collaboration with the IT team at NYUAD.

\subsection{Scientific software under the hood}
To fulfil its scientific tasks \f\ relies on open access or internal scientific software that is activated by the users via the front-end module.
The following is a brief description of the main components.

\begin{itemize}
\item \textbf{VOU-Blazars} \\ 
VOU-Blazars\footnote{\url{https://github.com/ecylchang/VOU_Blazars}}\citep{VOU-Blazars}, a tool developed as part of the Open Universe initiative, is the multi-frequency and multi-epoch data retrieval engine of \f. This software is used to produce multi-frequency sky maps suitable to locate blazars and blazar candidates in the error regions of X-ray, \gr,  high-energy neutrinos, and any other source with non-negligible localisation uncertainty. VOU-Blazars is also used to build Spectral Energy Distributions (SEDs) and light-curves. \\
\f\, currently uses VOU-Blazars V2.00, an improved version of the original code that provides access to an expanded list of over 90 catalogues and spectral databases using the protocols defined by the International Virtual Observatory Alliance. Some examples of recently added catalogues are eRosita-eFEDS, NEOWISE, unWISE, PACO, RATAN-600, GAIA, VLASS, RACS, NuSTAR, 4XMM-DR13 and Fermi 4LAC-DR3 and 4FGL-DR4. 

A more complete list of the catalogues, spectral and time-domain databases accessed by this version of VOU-Blazars is given in Table \ref{tabVOU}. 

\unskip
\begin{deluxetable}{ll}
\tablecaption{ List of the main catalogues and spectral databases queried by the version of VOU-Blazars currently implemented (V2.00) in \f.\label{tabVOU}}
\tablewidth{0pt}
\tablehead{
\colhead{\textbf{Energy band}	} &
\colhead{\textbf{Catalogues/databases}}
}
\startdata
Radio & NVSS, VLSSR, VLASS-QL, SUMSS, RACS, FIRST, TGSS, LoTSS, PMN,\\ & GB6, GB87,
 North20, WISH, GLEAM, ATPMN, K\"uhr, VLSS, AT20G, \\& ATPMN, CRATES, RATAN-600, PACO \\
$\mu$-waves/mm & ALMA, Planck:PCCS 39, 44, 70, 100, 143, 217, 353, 545, 857 \\
Infrared & IRAS, AKARI:BSC,PSC, WISE, unWISE, NEOWISE, 2MASS, DENIS,\\ & Herschel:SPIRE250,350,500,PACS70,ATLAS \\
Optical & USNO, HSTGSC, SDSS, PanSTARRS, Gaia, SMARTS\\
UV & GALEX, Swift-UVOT, XMM-OM\\
X-ray & Einstein:IPC, IPCSlew, EXOSAT:CMA, ROSAT:WGA,RASS, BeppoSAX, \\ 
& 4XMM-DR13, XMMSL2, Swift:2SXPS, OUSX, XRTSPEC, BMW, BAT105\\
&Chandra:CSC2, NuSTAR:NuBlazars, eROSITA:eFEDS,MAXI-GSC\\
\gr\ & Fermi:2FHL,3FHL,4FGL-DR4,2BIGB,MST12Y,1FLE,FMONLC,AGILE,\\ & VERITAS, MAGIC, HESS\\
Source catalogues & 5BZCat, 3HSP, 4LAC-DR3, ZWClusters, Abell, MilliQuas\\
& Pulsars, CVCat, SNRGREEN, MWSC, MWMC, XRBCAT
\enddata
\end{deluxetable}

\item \textbf{BLAST} \\
BLAST\footnote{\url{https://github.com/tkerscher/blast}}\citep{BLAST} is a machine learning tool that helps characterising the SEDs built within \f\ by estimating the position of the peak of the synchrotron emission (usually referred as \nup\ ), which is an important parameter for physical models and the classification of blazars.
The \nup\ values reported in the literature are often estimated manually and therefore their value can vary depending on the author and the method used. BLAST provides an objective estimation of \nup\  as well as an evaluation of its uncertainty based on the training of the tool with a large number of known blazars with good SEDs and \nup\ estimated by fitting the data to a polynomial function.

\item \textbf{W-Peak} \\
W-Peak \citep{Wpeak} is a blazar SED \nup\, estimator based on the average 4.6-3.5 $\mu$ spectral index as determined from WISE and NEOWISE infrared data. The infrared slope, when due to the non-thermal emission from the jet, is linearly related to the average \nup\, of blazars, as shown in Fig. \ref{irslopeVspeak} where 
the mean infrared slope is plotted as a function of \nup\ in the subsample of blazars frequently observed by Swift \citep{GiommiXRTspectra} whose IR flux is not due to the host galaxy, the infra-red torus or other components not related to the jet.
The best fit to the data shown in Fig. \ref{irslopeVspeak} is

\begin{equation}
Log(\nu_{\rm peak}) = 3.8\times \alpha_{(3.4-4.6\mu)}+13.9
\label{eq:1}
\end{equation}
Where $\alpha_{(3.4-4.6\mu)}$ is the spectral slope between 3.4 and 4.6$\mu$, calculated in the $\nu$f$_{\nu}$, vs $\nu$ space, averaged over all the available WISE and NEOWISE flux measurements. 

The W-Peak estimation, especially when using NEOWISE data, is based on infrared measurements averaged over a long period of time and it is therefore representative of the mean value of \nup\ of an object, and cannot be used to determine the peak value during flaring or low level states.  
As part of the process of estimating \nup\ the W-peak software also generates a table including the results of basic variability statistics and parameters such as variability amplitude, fractional variability, and minimum absolute deviation for NEOWISE data.
\begin{figure}
\includegraphics[width=16. cm]{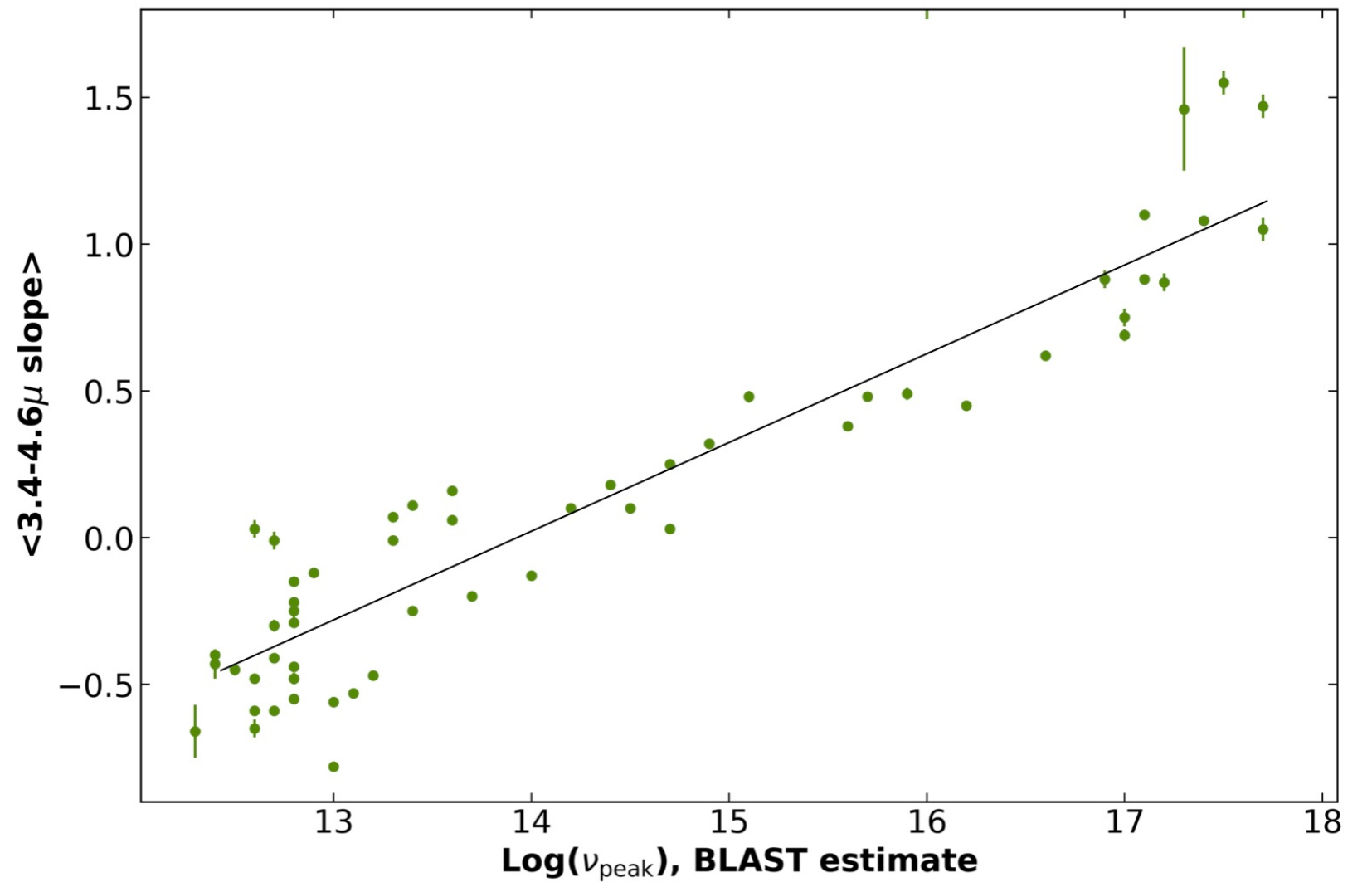}
\caption{The correlation between the average 3.4-4.6$\mu$ spectral slope and \nup\ in the subsample of blazars frequently observed by Swift whose IR can be attributed to the jet.
\label{irslopeVspeak}}
\end{figure} 
\item  \textbf{Aladin}\\
Aladin is a widely used powerful visualiser of astronomical images that was developed at CDS,  (\url{https://aladin.u-strasbg.fr}). The specific implementation of Aladin in \f\  provides access to a large selection of high quality surveys at radio, IR, Optical, X-ray and \gr\ energies, that are useful for blazar identification.  Table \ref{tabAladin} gives the list of the surveys that are available by default. 

\begin{deluxetable}{ll}
\tablecaption{List of the main surveys that can be accessed via the \f\ aladin interface\\ 
\label{tabAladin}
}
\tablewidth{0pt}
\tablehead{
\colhead{\textbf{Energy band}} &
\colhead{\textbf{Survey name}	}
}
\newcolumntype{C}{>{\centering\arraybackslash}X}
\startdata
radio & NVSS, VLASS (Epoch 1, 2.2 and 3.1), VCSS1, TGSS-ADR, SUMSS, RACS-low, RACS-mid \\
infrared & WISE, unWISE, 2MASS, Spitzer, Herschel \\
optical & DSS2, PanSTARRS, SDSS, ZTF, DECaLS, BASS, MAMA, SkyMapper, XMM-OM, Swift-UVOT\\
UV & Galex, Swift-UVOT \\
soft X-ray & RASS, Swift-XRT, XMM, Chandra  \\
hard X-ray & Swift-BAT, MAXI\\
\gr\ & Fermi-LAT \\
\enddata
\end{deluxetable}
\unskip
\end{itemize}

\subsection{Tables of known blazars, blazar candidates and other multi-wavelength emitters}
\f\, provides access to a catalogue of known blazars including over 6,400 objects, to tables of blazar candidates compiled in a variety of ways, and lists of other types of multi-wavelength emitters, as described in the following.

 \begin{itemize}
\item \textbf{The \fb\, reference list of blazars.}

This table combines the objects listed in the 5BZCAT \citep{Massaro2015} and the 3HSP \citep{3HSP} catalogues with the blazars and \gr-detected AGN of the Fermi 4LAC-DR3 catalogue \citep{4LAC-DR3}. 
This reference list is intended to be the starting point of a living blazar catalogue that will be periodically updated with newly published blazars and with the blazars that will be discovered with \f\, also via citizen researchers projects at NYUAD. 

To facilitate the discovery of new blazars  by \f\, users, in particular by citizen scientists or students, the next few tables provide lists of astronomical sources that have been selected with different criteria to match the broad-band spectral characteristics of blazars. They are expected to include 10-20\% of real blazars.
\\
\begin{itemize}
\item \textbf{Blazar candidates in the VLBI Radio Fundamental Catalog.} 

The radio fundamental catalog (RFC)\footnote{\url{http://astrogeo.org/rfc/}} is a compilation of VLBI-observed compact radio sources that is statistically complete above a flux density of 150 mJy at 8 GHz. 
 The \f\, table of blazar candidates  consists of all the sources in the RFC complete subsample that a) are positionally coincident with one of the X-ray sources of the RASS, Swift-XRT, XMM or Chandra catalogs, b) have a radio to X-ray flux ratio that is within the range observed in blazars and c) are not included in our reference list of blazars. 
\\

\item \textbf{Blazar candidates in the eRosita eFEDS survey.}
The Final Equatorial Depth Survey (eFEDS) is a soft X-ray (0.2-8 keV) survey of 140 square degrees of sky carried out during the performance verification phase of the SGR/eRosita satellite \citep{erosita}. The catalogue of detected sources in this region \citep{efeds} includes 32,684 objects. Our list of eFEDS blazar candidates consists in the subset of eFEDS detections that match the position of a radio source with flat-spectrum ($\alpha_{\rm r} > -0.7$)  estimated using the 1.4 GHz NVSS and the 3.0 GHz flux densities of the NVSS and the VLSS catalogs.

The list of eRosita blazar candidates will be largely expanded in the near future when the catalogues of eRosita surveys of large parts of the sky will be published.
\\
\item\textbf{Blazar candidates among the Swift-XRT serendipitous sources.}
The Neil Gehrels Swift observatory \citep{swift} has been observing the X-ray sky since 2004. A recent analysis of entire Swift X-Ray Telescope (XRT) archive led to the creation of a catalogue of $\sim$ 150,000 soft X-ray sources (Zazza et al. 2023), many of which are not associated with known astronomical objects.
The \f\, table of Swift-XRT blazars candidates consists of the subset of still unidentified X-ray detections that match the position of a radio source.
To limit the table to a manageable size and to increase the fraction of real blazars that could be detected in high-energy surveys (that is blazars with large \nup) we chose to select sources with (0.5-10 keV) X-ray to radio flux ratio larger than $5\times 10^{-11}$  erg/cm2/s/Jy. These selection criteria may change in the future to provide larger samples.
\\

\item \textbf{Blazar candidates in Fermi-LAT unidentified \gr\, sources.}
Most of the counterparts of the Fermi-LAT \gr\, sources located away from the Galactic plane are identified with blazars. However, a significant fraction still remains unidentified, although it is likely that a good percentage of these still unassociated sources will eventually be identified with Blazars. The list of still unassociated Fermi-LAT sources provided in \f\, consists of all sources in the 4FGL-DR4 catalogue \citep{4FGL-DR3} located at Galactic latitudes $|b| > 10^{\circ}$ and is expected to include several \gr\, emitting blazars that could be discovered and characterised by \f\, users. 
\\
\item\textbf{Blazars and blazar candidates in the error regions of IceCube neutrino tracks.}
The detection of a flow of high-energy astrophysical neutrinos by the IceCube South Pole observatory \footnote{\url{https://icecube.wisc.edu/}}
opened a new window on the Universe \citep{IceCube2013}. 
A number of papers reported evidence that Blazars are likely responsible for generating at least a fraction of the astrophysical neutrino flux that has been detected by IceCube \citep[][and references therein]{Aartsen2018,2021Univ....7..492G,pks0735}.

Over the last decade the IceCube team published a number of high-energy neutrino tracks, as alerts communicated via GCNs, immediately after their detections. These events have a good probability of being of astrophysical origin and have relatively small uncertainty regions ($\approx$ 1-4 degrees in size), although still significantly larger than those of electromagnetic sources. 
Very recently the IceCube collaboration published a new list of likely astrophysical neutrino track-like events detected between 2011 and the end of 2020 \citep{IceCat1}. 
The table provided in \f\, lists all the 274 tracks of the \citep{IceCat1} catalog as well all the neutrino tracks
published via the GCN channel from 2021 to the time of writing. The \f\, list currently includes over 330 events and will be kept up to date adding new events shortly after they are announced.\\
\end{itemize}

\item\textbf{Users provided lists of blazar candidates.}
Firmamento users can provide their own list of sources by uploading a simple "comma-separated values" (csv) file, which 
must include the source name, ra and dec (in degrees), and (optionally) parameters defining  the search radius around the source and the parameters of the uncertainty region, which is assumed to be of an elliptical shape. A comment 
field can also be added. The following is an example of the required format:

\begin{verbatim}
      Name,ra,dec,fov,major,minor,angle,comment
      source1,127.51611,1.57949,0.3,0.2,0.15,40.0, comment for source 1
      source2,139.48841,-1.54498,0.4,0.2,0.2,0.0, comment for source 2
      ...
      other sources details with the same format
\end{verbatim}
Where "fov, major, minor, angle" are the size of the area including the uncertainty region, 
and the parameters of the ellipse (in units of arcminutes and degrees).
If the file only includes the fields "name, ra and dec", that is fov, major, minor and angle are missing, then
firmamento assumes that the position of the source has no uncertainty and it will only generate the SED using the multi-frequency data that match the specified position.
\item \textbf{Other catalogs and tables of multi-frequency emitters.}

\f\, also provides tables and catalogs of known astronomical sources of a variety of types.

\begin{itemize}
    \item Catalogs of high-energy sources (e.g. Blazars in the TevCat, the Fermi 4FGL-DR4 and the 2Agile catalogs). 
    \item Selected tables of non-blazars multi-frequency sources.
    These are tables of astronomical objects that are in general not related to blazars, but that are nevertheless useful for general or educational purposes in the context of multi-messenger astrophysics.
\end{itemize}
\end{itemize}


\section{Firmamento for citizen scientists and students}
\subsection{The NYUAD Citizen Researches initiative and \f\ }
Citizen Researcher at the New York University-Abu Dhabi\footnote{https://citizenresearcher.hosting.nyu.edu/} is a research public participation initiative, developed by NYUAD that encourages members of the public in the UAE and elsewhere to participate and contribute to research in science, engineering, social science, arts and humanities. 
 The NYUAD Citizen Researcher initiative is a program that embodies the ethos of Citizen Science, a global movement where public involvement in scientific research goes beyond mere participation to encompass education, contribution, and collaboration. Unlike the conventional understanding of citizen science, which may simply involve data collection by the public, Citizen Researcher integrates participants into the heart of the research process. This program aims not just at assisting with data collection but at fostering a deep educational experience where participants develop research skills, engage in analysis, and contribute meaningfully to the outcomes of research projects across a wide spectrum of disciplines.
Most importantly, Citizen Researcher removes the barrier that research can only be done by professionals and in professional settings. Based on best practices from the field of Public Participation in Scientific Research, Citizen Researcher projects at NYUAD are designed to ensure that public participants can clearly and easily take part and have a meaningful learning experience where they can make an impact on critical research that has local and global significance. 
Citizen Researcher projects, such as Firmamento, transcend traditional public participation by not only providing tools and training for tasks like the astronomical sources validation process but also by leveraging the public's contribution in the context of high-level research, utilizing sophisticated methods like machine learning to enrich the data collected by participants. This innovative approach contributes to democratizing the research process, enabling anyone with interest and dedication to actively partake in research. Through this initiative, NYUAD makes citizen science a deeply immersive and educational experience that accelerates research outcomes and broadens the impact of academic studies by tapping into the collective effort and wisdom of the community.

\subsection{A catalog of new Blazar candidates with 
\f\, by High School students}
In 2022, four students from the Liceo Scientifico Statale Ugo Morin high school in Venice, Italy, participated to a citizen science program within the framework of the Italian MIUR PCTO programme, which translates to “Paths for 
cross-disciplinary skills and orientation”. The program aims at making students acquire skills outside the standard educational program. Within this agreement, the students invested around 40~h each on a program with the Department of Physics and Astronomy of the University of Padova under the supervision of researchers.
For this work the students started with \texttt{Fermi}-LAT data~\cite{Fermi-LAT}. In the 4FGL catalog (DR3)~\cite{4FGL} there are 6658 sources out of which several hundreds are blazars. Out of all the unassociated sources in this catalogue, a selection was done based on spectral hardness and distance from the galactic plane. The students were given a list of 198 unidentified LAT sources as input, selected by spectral hardness and location. Their goal was to find counterparts at other wavelengths and eventually propose an identification.

The first step was to verify whether these sources had counterparts in any other wavelength. This check was performed with   
\f\, inserting the coordinates of the region and of the parameters of the uncertainty area. In a second step, the students verified each single candidate association one by one. This was done by inserting the candidate coordinates again in 
\f\, and evaluating the  skymap in the optical spectrum around the direction of the candidate. In a third step, the students generated the Spectral Energy Distribution (SED). As a result, 54 new blazar candidates were found, characterized by the synchrotron peak with the BLAST code~\cite{BLAST} and by redshift where existing~\citep{Fronte:2022ite}. The candidates were labelled as with LLSUM acronym following the name of the high school. An extract of the candidate table is shown in \autoref{tab:lssum}.
\begin{table}[h!t]
  \centering
  \begin{scriptsize}
  \begin{tabular}{lcc|lcccc}
  \hline\hline
\texttt{Fermi}-LAT ID                &  RA & Dec & LSSUM ID & RA & Dec & $z$ & $\log_{10}\nu_{\rm{peak}}$ \\
\hline
4FGL J0000.7+2530	& 0.188	& 25.515	& LSSUM J000027.9+252805	& 0.11633& 25.4680	& 0.49	& 16.6	$\pm$0.5\\
4FGL J0026.1$-$0732	&6.540	&-7.543	    & LSSUM J002611.6$-$073115	&6.54842 &-7.52097	& -	&16.9	$\pm$0.4\\
4FGL J0045.8$-$1324	&11.472	&-13.403	& LSSUM J004602.8$-$132422	& 11.51154 & -13.4060&-	&15.6	$\pm$0.6\\
4FGL J0055.7+4507	&13.940	&45.124	    & LSSUM J005542.7+450701	&13.92792	&45.11706 &-	&15.8	$\pm$0.5\\
\ldots &&&&&&&\\
4FGL J1628.2+4642	&247.063&	46.715	& LSSUM J162755+464249	&246.98105 & 46.71342 &0.2135	&15.8	$\pm$0.4\\
4FGL J1658.5+4315	&254.646&	43.254	& LSSUM J165831.5+431615 &254.63126 & 43.27085	&0.63 (Phot)	&16.0	$\pm$0.5\\
4FGL J1706.4+6428	&256.606&	64.475	& LSSUM J170623.3+642725	&256.59688 & 64.45706	&0.27 (Phot) &15.9	$\pm$0.7\\
4FGL J1727.1+5955	&261.776&	59.926	& LSSUM J172640.4+595549	&261.66833 & 59.93036	&featureless&16.1	$\pm$0.4\\
\ldots &&&&&&&\\
\hline\hline
\end{tabular}
  \end{scriptsize}
\caption{\label{tab:lssum}An extract from the preliminary LSSUM (Liceo Scientifico Statale Ugo Morin) catalog of blazar candidates obtained with 
\f\,~\citep{Fronte:2022ite}. In the redshift column, "phot" means the redshift is taken from SDSS17 or NED and not from the galaxy spectra, while "featureless" is in case there are no optical lines.}
\end{table}

Plan to re-evaluate the targets by relaxing the criteria on the number of \texttt{Fermi}-LAT unassociated sample are being carried out as a follow-up project. Discussion about the possibility to carry out proposal of observation in optical (for redshift estimation) and in X-ray and gamma-ray to validate the inverse Compton peak are also being put forward. The four involved students remarked~\citep{Fronte:2022ite}:
\begin{quote}
\textit{
“This PCTO experience has been a fundamental opportunity to grow as persons, it gave us the possibility to see the research environment in close contact and to understand what working at a University really means. Thanks to this occasion we now know that we’d like to have a career as researchers someday.”}
\end{quote}

We plan some follow-up initiatives, such as submitting proposals for observations of sources with no redshift, and proposing similar undertakings to other students, both in Italy and at NYUAD.

\section{\f\, for astro and particle physicists}
In the following we give some examples of possible uses of \f\, in the framework of contemporary astro-particle research. 

\begin{figure}
\centering
\includegraphics[width=16.5 cm]{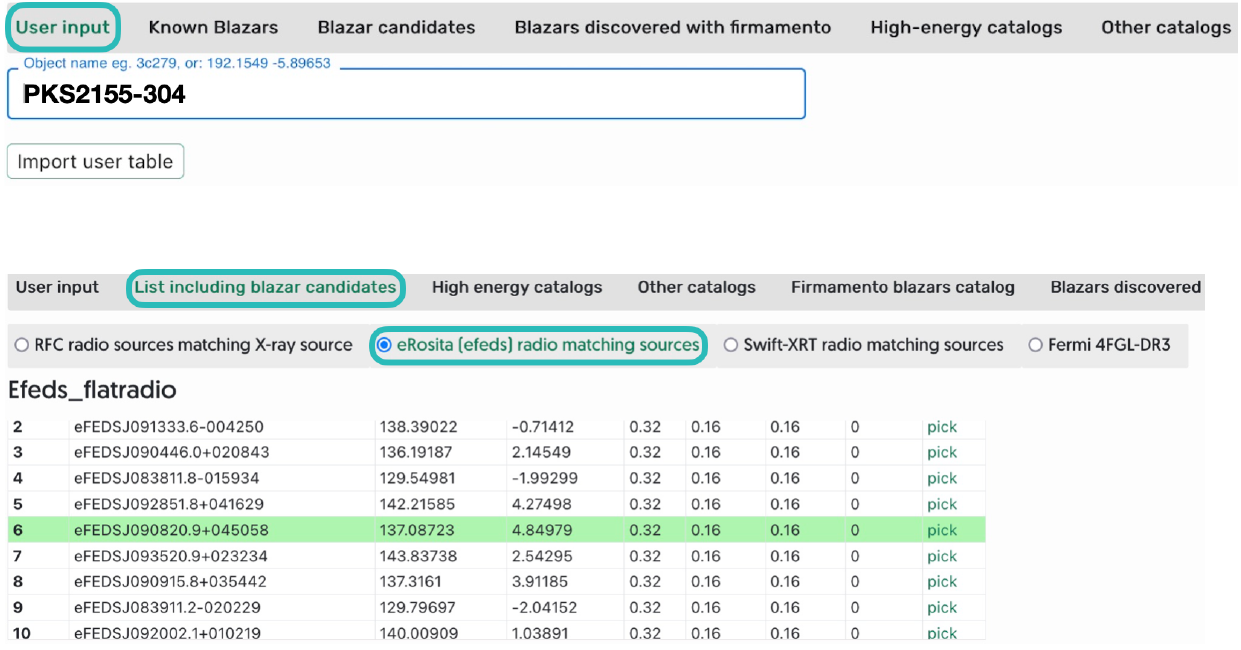}
\caption{The name or the position of an astronomical source, or of a localisation error to be processed  by \f\, can be provided either by specifying a source name, a set of RA, Dec coordinates or by selecting one entry from a list uploaded by the user (top), or by choosing one entry from one of the lists provided, e.g. the table of eROSITA detections from the eFEDS survey matching flat-spectrum radio sources, like in the lower part of the figure.\label{firmamento_samples}}
\end{figure}

\f\, provides multi-frequency data of astronomical objects whose common names or positions are entered by the users (like most web services providing astronomical data), or are included in lists assembled in various ways, e.g. from radio catalogues, X-ray and \gr\, detections or high-energy neutrinos, as described above. Fig \ref{firmamento_samples}
shows the part of \f\, where the user directly inputs the name of a source (top) or chooses one object from the 
table of eRosita sources (bottom).  

\subsection{The SED of a well-known and often-observed object: BL Lacertae}
\begin{figure}
\centering
\includegraphics[width=18.0 cm]{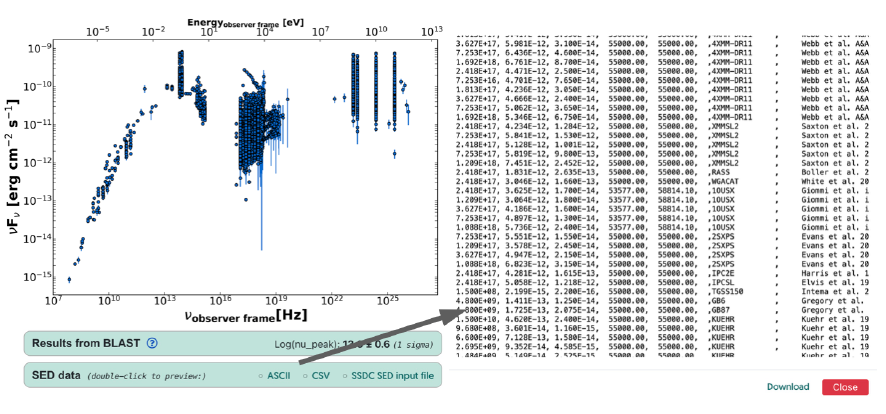}
\caption{The SED of the well-known source named BL Lacertae, the archetype of the class of blazars. The right side shows part of the corresponding data file, which can be downloaded by clicking next to the chosen format.\label{sed_bllac}}
\end{figure}

In this example we consider the SED of BL Lacertae (also known as BL Lac), the prototype of the class of blazars. This object has been observed with many ground-based and space observatories a large number of times over the years.  Fig. \ref{sed_bllac} shows the SED generated with \f\.. Note the very large density of data available in most energy bands and in particular in the X-ray region where the many spectral measurements from the systematic analysis of Swift \citep{GiommiXRTspectra} and NuSTAR \citep{middei2022} blazars observations are plotted.
 All SED data points, inclusive of observation time and corresponding bibliographic references, can be downloaded in different formats with a simple click, as shown in the figure. These data can be selected by the users for specific needs, e.g. fitting a model during a particular intensity state or time interval. In the future \f, will provide tools to access the data depending on specified time intervals or according to other selection criteria.
\subsection{Locating blazars candidates in unidentified FERMI 4FGL-DR3 \gr\, sources}
\begin{figure}[h!t]
\centering
\includegraphics[width=18.0 cm]{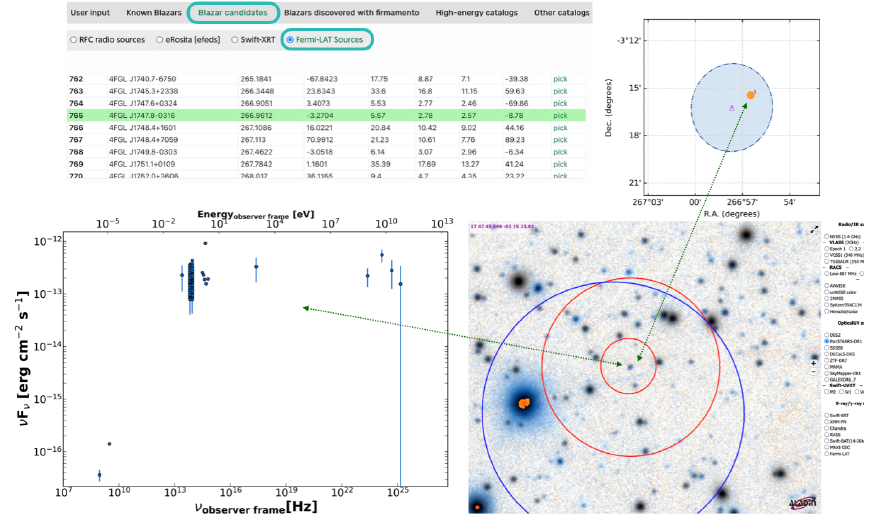}
\caption{Example of a possible identification of a blazar counterpart in the error ellipse of a still unidentified \gr\, source. Top left: the user selects a source from the list of unidentified Fermi 4FGL-DR3 detections. Top right: the 4FGL-DR3 localisation error ellipse of the chosen source with a possible blazar counterpart appearing as an orange symbol. Lower left: the SED of the blazar candidate. Lower right: the error regions of catalogued radio and X-ray sources (red and blue circles) matching the position of the candidate blazar are superposed to the image from a optical survey chosen in the \f\, aladin interface.  \label{FermiExample}}
\end{figure}

Here we show how \f\, can be useful to identify blazars in the error ellipses of Fermi-LAT sources. Fig \ref{FermiExample} shows the case of the \gr\, source 4FGL$~$J1747.8-0316, one of those listed in the table of Fermi unidentified sources (upper left); the error ellipse of this source generated with \f\, is shown in the upper right part of the figure.
The orange symbol is a candidate blazar selected by VOU-Blazars considering X-ray and radio matching sources. The bottom part of fig \ref{FermiExample} shows the SED of the candidate (left) and the error circles chart shown with Aladin using the PanSTARRS survey (image background) with overlaid the uncertainty region of the X-ray data (blue circle) and the radio measurement (red circle).

\subsection{The localisation error of IceCube200107A and the blazar 3HSP J095507.1+355100}
\begin{figure}[h!t]
\centering
\includegraphics[width=16.5 cm]{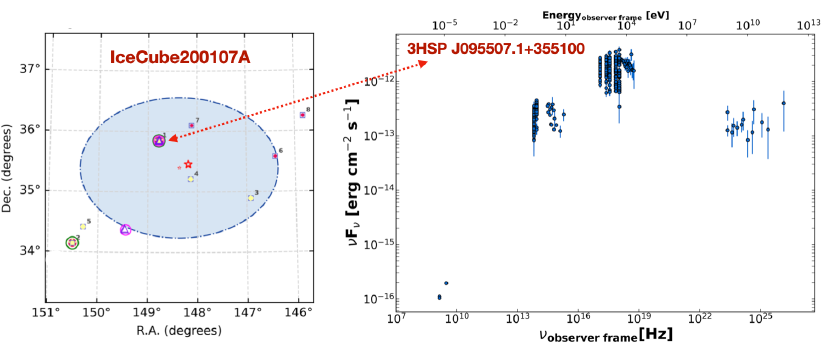}
\caption{The 90 percent localisation uncertainty of the IceCube neutrino track IceCube200107A and the SED of the possible electromagnetic counterpart 3HSP J095507.1+355100. See text for details.
\label{IC200107A}}
\end{figure}

The blazar 3HSP$~$J095507.1+355100 has been reported as the possible counterpart of the high-energy neutrino IceCube200107A \citep{2020A&A...640L...4G}. The left side of Fig. \ref{IC200107A} shows the 90\% localisation error of IceCube200107A (light blue inner area), which includes several candidate blazars, as generated by \f\, via VOU-Blazars. Only one of these sources, 3HSP J095507.1+355100, is also a \gr\, source, as marked by the open purple triangle. Its SED is shown on the right side of the figure. 
For more details on this case and about association of IceCube astrophysical neutrinos and blazars in general see \cite{2020A&A...640L...4G,Giommidissecting,Dissecting}.

\begin{figure}[h!t]
\centering
\includegraphics[width=16.0 cm]{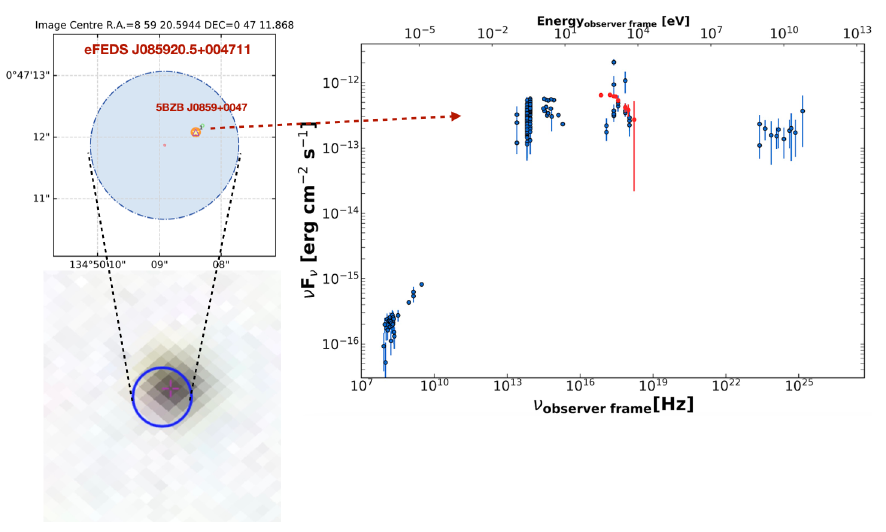}
\caption{Top left: the error circle of the source eFEDS J085920.5+004711 including the blazar 5BZB J0859+0047. On the lower left is shown the very small error circle overlaid to the SDSS optical image of this part of the sky. Right: The SED of the Blazar 5BZB J0859+0047 = eFEDS J085920.5+004711 generated with \f. The red points are from the eRosita eFEDS survey.\label{sed_erosita}}
\end{figure}

\begin{figure}[h!t]
\centering
\includegraphics[width=16.5 cm]{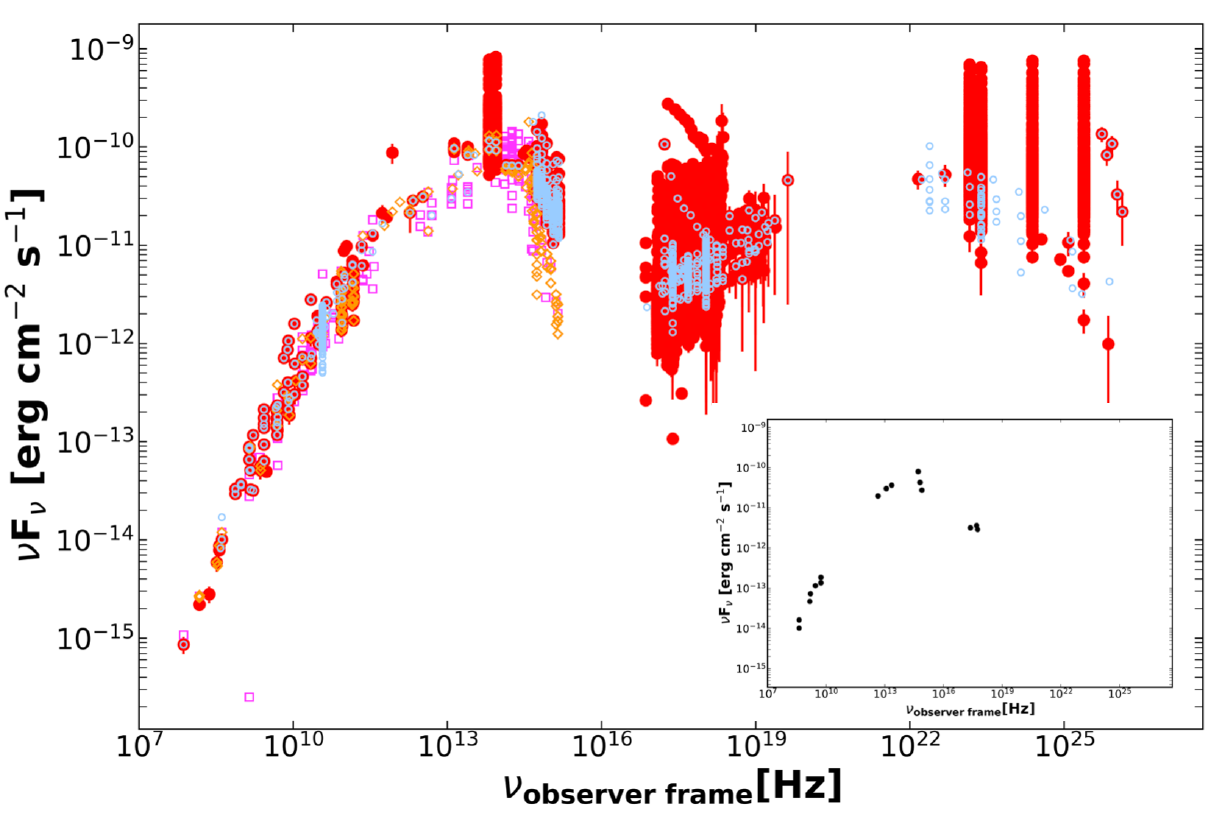}
\caption{A comparison of the SED of the prototype blazar source BL Lacertae obtained with Firmamento and other on-line SED tools. Firmamento data is shown as filled red points, SSDC SED builder as open light blue circles, NED as magenta open squares, and Vizier as orange open diamonds. The SED in the inset, adapted from \cite{giommiansarimicol}, illustrates the amount of multi-frequency data available from public sites in the mid 1990's.}  
\label{SEDcomparison}
\end{figure}

\subsection{The eRosita/eFEDS sample of X-ray sources with radio counterparts}

\f\, provides several lists of sources with associated positional uncertainties that potentially include blazars. As an example we mention here a sample of X-ray sources that are positionally consistent with radio sources, and have been detected in the eRosita Final Equatorial-Depth Survey \citep[eFEDS, ][]{efeds}, a dataset covering 140 square degrees of sky that was made publically available shortly after the launch of the mission. A specific case illustrating the detection of a blazar in eRosita data is shown in Fig. \ref{sed_erosita} where the X-ray error region of the source eFEDS~J085920.5+004711, which includes the blazar 5BZBJ0859+0047, as well as the corresponding SED, are shown. The eRosita X-ray data points, converted to \nufnu\ units from the integrated fluxes reported in the catalog, appear as red symbols. Additional samples of eRosita sources will be provided shortly after these will be released by the eRosita teams.

\begin{figure}[h!t]
\centering
\includegraphics[width=12.cm]{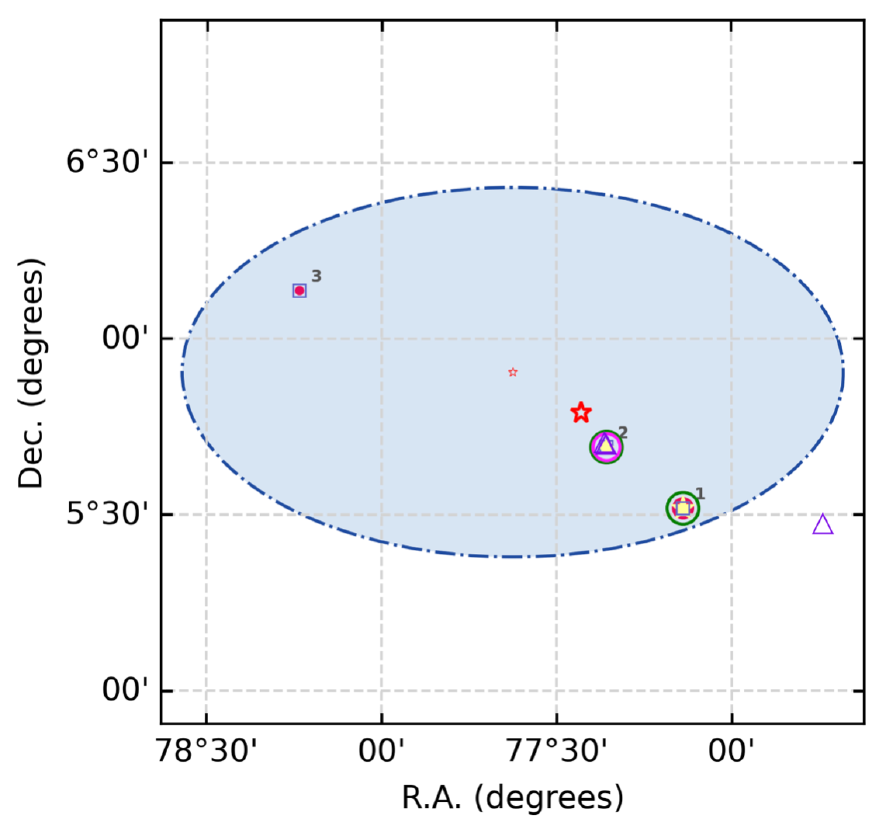}
\caption{The blue ellipse plottted in this figure illustrates the localization error region of IceCube170922. The only known source detected at \gr\, inside this area (identified by the number 2) is the Blazar TXS0506+056, which has been associated with the IceCube neutrino.} 
\label{IC170922}
\end{figure}

\section{Scientific validation and Comparison with other platforms}

As mentioned in the Introduction, for several years now a number of official archives have provided access to astronomical data generated by scientific satellites and ground-based observatories funded by agencies around the world, serving the international astronomical community. Alongside these institutional services a wealth of other websites offer public data and tools to the general astronomical community in many ways. Firmamento was conceived to follow a somewhat different concept and therefore distinguishes itself from other online platforms for several notable reasons:
\begin{itemize}
    \item 
    Specialization: Firmamento is a scientific topic-oriented tool with a current clear specialization in the field of Blazars and multi-messenger emitters. It places a strong emphasis on addressing contemporary scientific challenges, such as spectral and time-domain detailed studies and the identification of new blazars and other multi-frequency emitters in the error regions of high-energy sources, including X-ray, \gr\, and astrophysical neutrinos.
\item
    Comprehensive Data: The platform offers a comprehensive range of products in the energy, time and imaging domains. These include well-populated Spectral Energy Distributions (SEDs), time-domain data in several energy bands (e.g. IR, optical, X-ray and \gr\.), error regions maps with candidate counterparts, visualisation of multi-wavelength surveys, lists of candidate blazars, catalogs of high-energy sources, etc. 
\item
    Value-Added Information and data characterization: By incorporating value-added information generated through machine learning  and other techniques  \f\, seeks to help characterizing the data retrieved and therefore enhance the quality and depth of the services it provides. Examples are the BLAST and W-peak, tools that estimate of the position and intensity of the synchrotron peak of the SED, two important observational parameters in contemporary blazar research.  
\item
    Inclusivity: Building on the principles of the Open Universe initiative, one of Firmamento's core objectives is to provide quality services that cater not only to professional scientists but also to citizen scientists and the educational sector. This inclusive approach ensures that its resources are accessible and beneficial to a broader audience.
\end{itemize}
To scientifically validate \f, and assess its effectiveness, in the following we compare some of the data products generated using \f\, with those produced by other on-line platforms or have been published in the scientific literature.

 Figure \ref{SEDcomparison} presents a comparison of the SED for the Blazar BL Lacertae built using \f\, with those produced using other SED analysis tools accessible online. Firmamento data points are shown as filled red circles, while those from the SSDC SED builder \footnote{\url{https://tools.ssdc.asi.it/SED}} appear as overlaid open light blue circles, those from NED \footnote{\url{https://ned.ipac.caltech.edu}} as magenta open squares, and those from the Vizier photometry viewer \footnote{\url{http://vizier.cds.unistra.fr/vizier/sed/}} as orange open diamonds. Note the large overlap and agreement among all the data sets in the region $10^{8}$Hz to $\sim 3\times 10^{14}$Hz. At higher frequencies, up to a few times $10^{15}$Hz, the NED and Vizier points are systematically lower than those of Firmamento and of SSDC. That is because NED and Vizier provide flux values "as observed" without any correction to compensate for absorption in our Galaxy. These points should be corrected before they can be compared to theoretical emission models. The much larger amount of \f\, data points in the X-ray bands is due to the inclusion of the spectral data from many Swift and NuSTAR observations that have been analysed and made publically available in the framework of the Open Universe initiative \citep{GiommiXRTspectra,middei2022}.
In the inset of Fig. \ref{SEDcomparison}, we showcase an early SED of the same blazar, as initially presented by \cite{giommiansarimicol} in the mid-1990s using data from the European Space Information System \citep[ESIS,][]{esis}, one of the first on-line sites to serve multi-frequency data at a time when multi-wavelength astrophysics was still in its early stages. The large disparity in data abundance (18 vs approximately 17,700 flux measurements) and the large variability observed reflects both the recent exponential growth in available astronomical data and the close interconnection between multi-frequency and time-domain astronomy.

One of the most distinctive features of \f\, is its ability to facilitate the identification of blazars in the localization uncertainty regions of high-energy sources.
To use a well-known case to compare the maps generated by \f\, with those that appeared in the literature, Figure \ref{IC170922} shows the error region of the famous astrophysical neutrino IceCube170922 which has been associated the blazar TXS0506+056. 
This figure is very similar and can be directly compared with e.g. Fig. 2 of \cite{IceCube170922Science} and Fig. 1 of  \cite{Dissecting}, the papers that identified TXS0506+056 as the first neutrino source.

\section{A mobile-friendly tool}
\begin{figure}[h!t]
\centering
\includegraphics[width=16. cm]{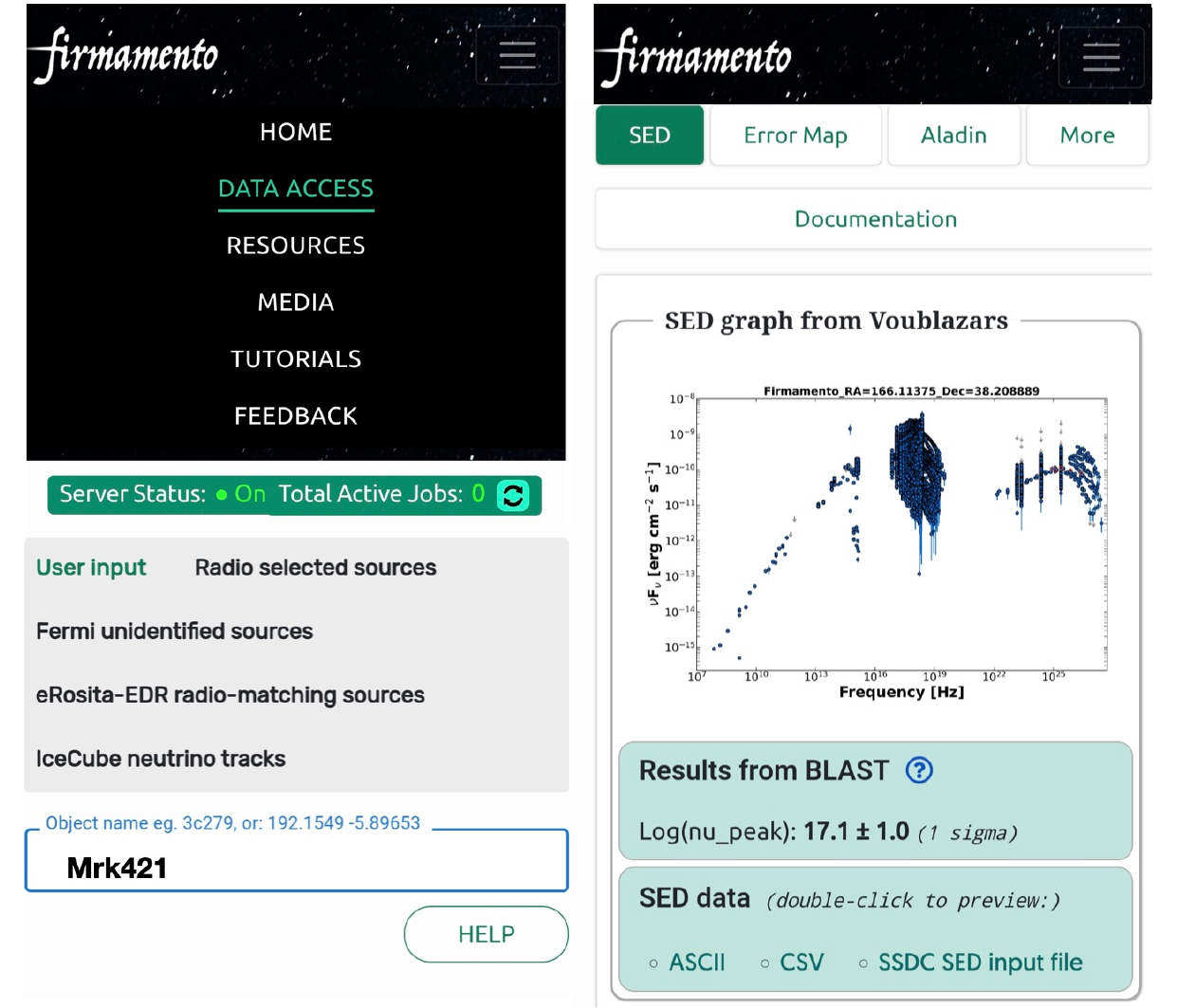}
\caption{Screenshots of \f\ running on a mobile phone. Left: the input area, where the user can enter a source name or choose from a list of 
radio, X-ray, \gr\ or IceCube neutrino detections. The right part shows the SED of the blazar MRK421, together with the estimation of \nup\, from BLAST and the possibility to download the SED data in different formats. \label{mobile}}
\end{figure} 
In an effort to offer a good degree of flexibility and usability by the widest possible range of users with different skills and preferences, including the participants to the citizen researcher project at NYUAD, \f\ has been designed to be a tool that is both mobile and computer friendly.\\
To illustrate how \f\, can be used on a smartphone Fig. \ref{mobile} shows two screenshots corresponding to a request for the SED of a well known Blazar. On the left side the user, after choosing the "Data Access" option, types the name of the requested source (Mrk421) in the input area, and then clicks on the "Run" button. The right side shows the result with a picture of the requested SED generated using VOU-Blazars, the result from the BLAST tool, and various options to continue, including the downloading of the SED data in different formats. \\

\section{Conclusions and future perspectives}
We presented \f\, a web-based and mobile-friendly tool dedicated to blazars and multi-frequency/multi-messenger emitters in general. 
Since many aspects of modern astrophysics rely on multi-frequency data, this new facility has the potential to become a valuable service supporting a broad range of topics in contemporary astronomy.  
\f\, has been designed to assist both professional and citizen scientists, with the goal of increasing the discovery power in the scientific community and of offering new opportunities for people with different skills to contribute to scientific advancement. We will use \f\, to support citizen researcher projects and we hope this system will be used in the scientific community.
\f\, can be accessed from all devices, including mobile phones, thus providing effortless access to a vast pool of science-ready, high-quality data and efficient tools for handling multi-frequency data. 
\f\, also employs advanced algorithms and machine learning tools to assist professional scientists in a novel way and makes astrophysical research more accessible to a wide range of individuals with varying levels of expertise.
This approach is expected to become widely adopted in the coming years as artificial intelligence technologies are integrated into online scientific services.

Along these expectations future versions of \f\, will expand its capabilities in different directions, starting with increasing the amount of data available by providing interfaces to other similar web sites, e.g. NED, Zwicky Transient Facility (ZTF)
\footnote{\url{https://www.ztf.caltech.edu/}}, Vizier etc. From the data analysis viewpoint we plan to offer more support to time domain analysis, which at the moment is limited to the timing parameters provided by the W-peak tool in the infrared band. 
To expand the potential for the discovery of new blazars, we will define lists of candidate blazars in the eROSITA all sky surveys which are expected to include several thousand new BL Lacs and FSRQs that could be discovered by \f\, users, hopefully with a good participation of citizen scientists and students.
In an effort to broaden support for multi-frequency emitters of all types, we plan to include methods to identify radio-quiet extragalactic sources, such as normal AGN and X-ray emitting stars, and to incorporate catalogs of Galactic sources. Subsequently, we aim to introduce algorithms suitable for discovering new Galactic sources based on their specific multi-frequency emission.
Finally, to expand the involvement, to increase the analysis capability and to stimulate the capacity to conceive and perform independent scientific projects by citizen scientists and to better serve the education sector we will interface \f\, with artificial intelligence tools, starting with ChatGPT or similar tools. These language generation tools undoubtedly have significant potential to enhance Firmamento’s services, beginning with a more comprehensive presentation of the results. Predicting the exact direction of the application of these methods is challenging; therefore, we plan to start gradually by training these tools with scientific papers related to the data provided by Firmamento. This approach will enable the tools to assist users, especially citizen scientists, in answering questions and understanding the physical models used to interpret the data.

\begin{acknowledgments}
\section{acknowledgments}
We thank the anonymous referee for useful comments that helped us improving the paper.
We thank the IT support at NYUAD for their help in the setup of the Firmamento server at NYUAD. 
Paolo Giommi expresses his gratitude to the Center for Astro, Particle and Planetary Physics (CAP3) for supporting his research visit at NYU Abu Dhabi. Narek Sahakyan acknowledges the support by the Science Committee of RA, in the frames of the research project No 23LCG-1C004.
This material is based upon work supported by Tamkeen under the NYU Abu Dhabi Research Institute grant CASS
\end{acknowledgments}
%

\vspace{5mm}


\software{VOU-Blazars \citep{VOU-Blazars},Aladin \citep{aladin}, BLAST\citep{BLAST}
 }




\bibliography{myreferences.bib}
\bibliographystyle{aasjournal}



\end{document}